# FRII Radio Sources in Rich Clusters of Galaxies

By Lin Wan & Ruth A. Daly

Department of Physics, Princeton University, Princeton, NJ 08544

A sample of FRII radio sources in rich clusters of galaxies, both at high (z~0.5) and low (z~0) redshift, has been constructed to study the effect of environment on radio sources. Comparisons are made between the properties of FRII sources in cluster and non-cluster environments, and between X-ray clusters with and without FRII sources. The principal results are the following:

1. Most low-redshift FRII sources in clusters appear to be similar to FRII sources in group or field environments in terms of radio power, optical properties of the host galaxy, and nonthermal pressure of the radio bridge. Cygnus A is an exception with its high radio power and high nonthermal pressure. Most low-redshift clusters with FRII sources tend to lie at the lower end of cluster X-ray luminosity distribution, some having $L_x$ comparable to FRII sources in non-cluster environments. However, there are exceptions such as Cygnus A. Together, these results are in agreement with the model that it is the high-pressured intracluster medium that prevents FRII sources from forming in rich cluster environments at low-redshift.

2. High-redshift FRII sources are all quite similar to each other irrespective of their environments. While the radio powers and emission line luminosities of the host galaxies are higher on average than low-redshift FRII sources, there is also overlap, and the nonthermal pressures of the radio bridges appear to be similar to those of low-redshift FRII sources. Unfortunately, the X-ray data for high-redshift clusters with FRII sources are inconclusive because of the large number of upper bounds involved and possible AGN contribution to the X-ray luminosity.

3. The nonthermal pressures of the bridges of FRII sources appear to be similar to the thermal pressures of the ICM around them. This result, if confirmed by a larger sample, would allow FRII sources to be used as probes of their gaseous environments. The fact that most high-redshift FRIIs have similar nonthermal pressures to their low-redshift counterparts indicates that the gaseous environments around them are also similar, suggesting that most high-redshift clusters with FRII sources should be under-luminous X-ray emitters as are their low-redshift counterparts. Thus, the evolution in the clustering strength around FRII sources toward high-redshift is likely to be closely linked to an evolution of the state of the intracluster medium.

## 1. Introduction

The effect of the environment on radio sources has always been an interesting problem and has been considered in many studies. It has been reported that at low redshift, FRI radio sources usually inhabit moderately rich cluster environments, while FRII sources tend to lie in either small groups of sub-Abell richness or isolated fields [1,2,3,4]. One of the proposed explanations for this difference between the environments of FRI and FRII sources is that rich clusters contain a hot intracluster medium (ICM) whose high pressure severely disrupts the jets coming from the radio core and/or the related shocks, and thus prevents the formation of the "classical-double" FRII sources, except for sources with very high beam power, such as Cygnus A [2,3,4]. However, as was pointed out by Prestage and Peacock in 1988 [2], there are rare yet important exceptions, namely, there are indeed FRII sources in rich clusters of galaxies at low-redshift, some of which are of moderate radio power as opposed to the very high radio power of Cygnus A.

More recent investigations of the cluster environments of radio sources at z∼0.5 reveal that FRII sources at that epoch tend to inhabit richer environments than at present, while the FRI sources show no change in environment between the two epochs [3,4,5].

If it is the high-pressured ICM that prevents FRII sources from forming in rich clusters at low redshift, then the change in the environments of FRII sources at $z \sim 0.5$ indicates a change of the ICM, a change of the power or pressure of FRII sources, or a combination. In the evolving ICM model, the ICM of some optically rich clusters at $z \sim 0.5$ might have lower pressures than their low-redshift counterparts and thus sustain an FRII source. In the radio power evolution model, high-redshift radio sources have higher radio power than low-redshift ones, and thus can survive as FRII sources in gas-rich environments. Hill and Lilly (1991)[4] suggest that the evolving ICM model might be important since: i) there is no dependence of radio power on cluster environment at z∼0.5, as was confirmed by the results in Allington-Smith 1993 [5]; and ii) the 1Jy and 5C12 sources at z∼0.5 in their sample have the same power as sources at low redshift, yet they inhabit much richer environments than the low-redshift sources.

Two approaches are taken to investigate whether the high-pressure of the ICM in rich clusters is the reason why FRII sources tend to avoid rich environments and whether the evolution in the clustering strength around FRII sources is caused by an evolution of the state of the ICM. The first is to compare the properties of FRII sources in cluster and non-cluster environments, both at high and low redshift. The second is to compare clusters containing FRII sources with clusters without FRII sources, again, both at high and low redshift. Since it is the gaseous rather than the galactic environment that is assumed to affect the morphology of the radio source, the best parameter to compare between clusters with and without FRII sources is their X-ray luminosities. If the ICM in clusters containing FRII sources are of relatively low pressure, then they will have relatively low X-ray luminosities.

With this in mind, we searched the literature and compiled a sample of 47

low-redshift (z<0.35) FRII sources in non-cluster environments, 16 low-redshift FRII sources in rich clusters, 20 high-redshift (0.35<z<0.7) FRII sources in non-cluster environments, and 18 high-redshift FRII sources in rich clusters. The radio properties of these FRII sources, as well as the optical properties of their host galaxies were compared. The X-ray properties of the clusters around FRII sources, wherever available, were also closely compared to different samples of X-ray clusters without FRII sources, both at high and low redshift. Values of the Hubble constant $H_o = 100$ km s$^{-1}$Mpc$^{-1}$ and the deaccelaration parameter $q_o$=0 are adopted throughout.

A detailed description of our samples, data, and analysis are presented by Wan & Daly 1995 [6]. The major results of this study are presented here, and preliminary results related to this study are presented by Wan *et al.* (1994) [7]. The major references for the data used are [8,9,10,11,12,13,14,15,16,17,18,19,22,23,24].

## 2. FRII sources in cluster and noncluster environments

### 2.1. *Comparison of radio power and optical properties*

Wan & Daly (1995) [6] compare the radio powers at 408MHz ($L_{408}$), the narrow emission line luminosities ($L_{em}$), and the absolute V magnitudes ($M_v$) of host galaxies of FRII sources in cluster and non-cluster environments, both at high and low redshift. Within each redshift bin and irrespective of galaxy environment, the FRII sources studied have similar $L_{408}$, $L_{em}$, and $M_v$, with the exception of Cygnus A, which has a radio power that is much higher than other low-redshift FRIIs. High-redshift FRIIs have higher $L_{408}$ and $L_{em}$ than low-redshift FRIIs on average. But there are also overlap regions. The magnitude of the host galaxy is similar for the high and low redshift FRIIs studied.

### 2.2. *Nonthermal and thermal pressure*

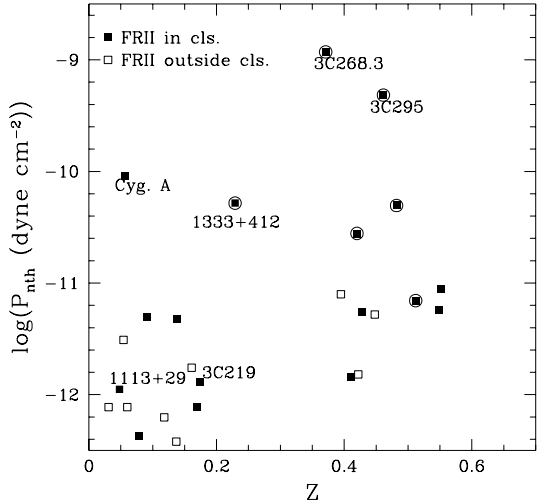 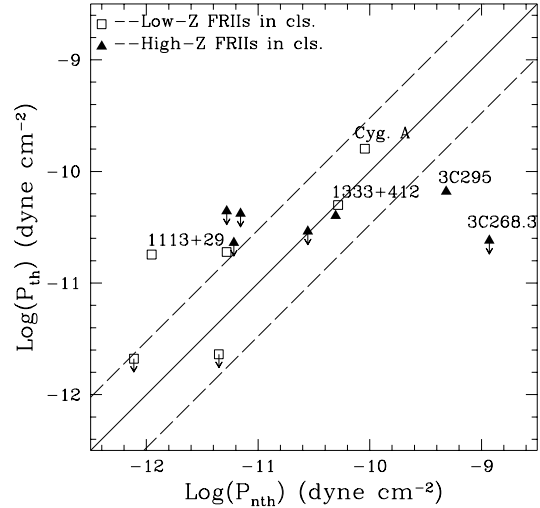

Figure 1  Figure 2

An important parameter of an FRII source is the nonthermal pressure of the radio bridge. The minimum-energy magnetic field strength ($B_{min}$) and nonthermal pressure ($P_{nth}$) of the radio bridge are calculated when a high resolution map of the FRII source is available (see [6] for details). In a given source, it is found that the B field strength stays roughly constant along the bridge, with a typical variation of ∼30%. For sources without high-resolution bridge maps, the average B field of the source is calculated. This average value is generally larger than $B_{bridge}$ due to the hot spots, which do not contribute to the bridge pressure. For the FRIIs whose $B_{bridge}$ is available, $B_{average}$ is usually about 1.5 times larger than $B_{bridge}$, though there is significant scatter about this value.

Figure 1 is a log-linear plot of $P_{nth}$ of the FRII sources versus redshift. Sources with only $B_{average}$ available are circled. It can be seen that most low-redshift FRIIs have similar $P_{nth}$ irrespective of whether they are in a rich cluster environment, though there are exceptions, such as Cygnus A. Similarly, high-redshift FRIIs also have $P_{nth}$ that appears to be independent of whether the source is in a rich cluster environment. Further, the high and low-redshift FRIIs have similar $P_{nth}$, which becomes obvious if only the sources with $B_{bridge}$ available are considered. The origin of the two exceptions, 3C268.3 and 3C295, is understood, and is discussed by Wan & Daly (1995).

FRII sources can used as probes of their environments if the nonthermal pressure of the radio bridge is in equilibrium with the thermal pressure of the surrounding medium. The facts that the bridges of many FRII sources do not undergo large amounts of expansion [25] and that the B field strength stays roughly constant along these bridges, tend to suggest that rough equilibrium between internal and external pressures has been reached.

In figure 2, the thermal pressure of the ICM around FRII sources estimated using X-ray data versus the non-thermal pressure of the radio bridges is plotted. The solid line indicates where $P_{th} = P_{n-th}$, and the dashed lines correspond to where $P_{th}$ is within a factor of 3 of $P_{n-th}$. Errors associated with the thermal and nonthermal pressures are rather large. Thus, $P_{th}$ and $P_B$ are taken to be roughly equal if they are within a factor of 3 of each other. It can be seen that most of the points fall into, or point toward, the "equilibrium" zone. It is perhaps not surprising to see that the two small sources, 3C268.3 and 3C295, have $P_{n-th}$ that are much higher than $P_{th}$. As discussed by Wan & Daly (1995), these sources are small and young, so there may not have been time for pressure equilibrium to be reached. Further, these sources are likely to be interacting with the interstellar medium. Thus, pressure equilibrium with the ICM is not expected.

### 2.3. *Conclusions*

This study of FRII sources in cluster and non-cluster environments suggests the following. 1) There appears to be rough equilibrium between the nonthermal pressure of a given radio bridge and the thermal pressure of the surrounding

ICM. Thus, FRII sources may be used to probe their gaseous environments. 2) At low-redshift, most FRII sources in rich clusters are similar to those in non-cluster environments in almost every respect. In particular, their nonthermal pressures appear to be similar to those of FRIIs in non-cluster environments. Thus, it is expected that their gaseous environments are of low pressure, as will be shown to be the case in §3 using X-ray data. 3) The radio and optical properties of high-redshift FRII sources in cluster and non-cluster environments are indistinguishable. High-redshift FRIIs have higher radio and emission line luminosities than low-redshift FRIIs on average, but their nonthermal pressures appear to be similar to those of low-redshift FRIIs, suggesting that the gaseous environments around them are similar to those around low-redshift FRIIs.

## 3. Clusters with and without FRII sources

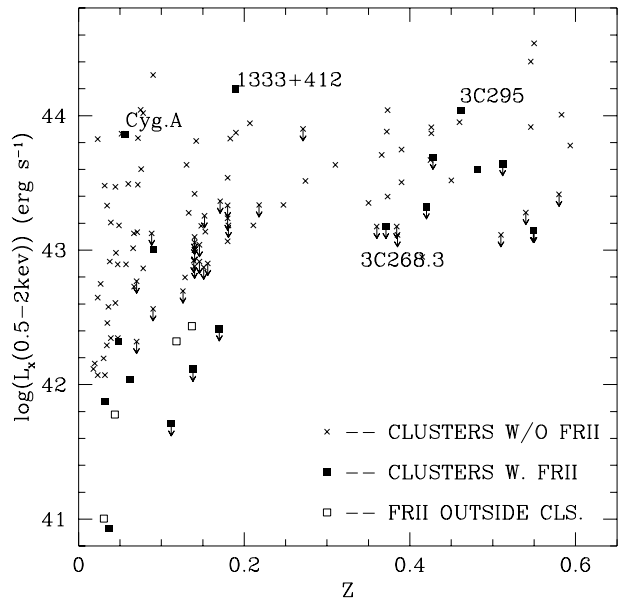

Figure 3

X-ray luminosities of all clusters with and without FRII sources versus their redshifts are plotted in figure 3; solid symbols represent clusters known to contain FRII sources. The low-redshift (z<0.35) comparison sample of clusters without FRII sources is taken from Abramopoulos and Ku (1983)[8] (hereafter AK83), which includes clusters with richness ranging from A0 to A3, and are shown as crosses in figure 3.

At low redshift, most clusters with FRII sources appear to lie at the lower end of the cluster X-ray luminosity distribution, suggesting that the ICM in these clusters has low pressure. The cluster A1763 around the radio source 1333+412 has a relatively large core radius of 270 $h^{-1}$kpc, so its high X-ray luminosity is probably a result of the large X-ray emitting volume rather than a high gas pressure. The X-ray luminosities of 4 low-redshift FRII sources in noncluster environments are plotted as open symbols in figure 3. It is interesting to note that some of the clusters with FRII sources have X-ray luminosities that are comparable to those in non-cluster environments.

Some numerical results follow. The average X-ray luminosity ($L_{x,ave.}$) for the 10 low-redshift clusters with FRII sources, including 3 upper bounds as detections, is $L_{x,ave.} = (2.49 \pm 1.63) \times 10^{43}$ erg s$^{-1}$, which drops to $(0.23 \pm 0.11) \times 10^{43}$ erg s$^{-1}$ if the two highest luminosity sources, Cygnus A and 1333+412, are excluded. The median X-ray luminosity is $L_{x,med.} = 0.17 \times 10^{43}$ erg s$^{-1}$, and is a more accurate measure of the typical X-ray luminosity of these clusters. This is to be compared with the average X-ray luminosity for the 72 low-redshift Abell clusters without FRII sources. Using AK83, including 20 upper bounds as detections, $L_{x,ave.} = (2.55 \pm 0.39) \times 10^{43}$ erg s$^{-1}$, and $L_{x,med.} = 1.33 \times 10^{43}$ erg s$^{-1}$.

More recent X-ray observations of low-redshift (0.03<z<0.15) Abell clusters by Burg *et al.* [26] indicate that $L_{x,med.} = 0.37 \times 10^{43}$ erg s$^{-1}$ for clusters of Abell class 0 (A0), $L_{x,med.} = 0.71 \times 10^{43}$ erg s$^{-1}$ for A1 clusters, and $L_{x,med.} = 4.03 \times 10^{43}$ erg s$^{-1}$ for A2 clusters. Noting that the clusters with FRII sources in our sample have richness classes ranging from A0 to A3, yet have a median X-ray luminosity which is lower than that of A0 clusters, it is clear that these clusters are indeed underluminous in the X-ray.

The high-redshift X-ray data are inconclusive because most of the clusters with FRII sources only have upper bounds on their X-ray luminosities as opposed to detections. Also, AGN contributions to the observered X-ray luminosities may be significant (see [6,13] for details).

The results presented in §2.3 suggest that high-redshift clusters with FRII sources should have gaseous states similar to low-redshift clusters with FRII sources. Thus, these clusters are expected to be underluminous X-ray sources and the X-ray luminosities of the ICM of these clusters are expected to be much lower than the current bounds.

## 4. Discussion

The results of this study show that most FRII sources in clusters are similar to those in noncluster environments, and that most of the clusters around these FRII sources are underluminous in the X-ray compared to clusters without FRII sources. This suggests that the ICM in these clusters is of low pressure. These results are consistent with the hypothesis that FRII sources tend to avoid rich clusters because of the high pressure of the ICM.

It is unfortunate that, given the present high-redshift X-ray data, direct estimates of the properties of the ICM in high-redshift clusters with FRII sources can not be assessed. However, utilizing the nonthermal pressures of these FRII sources as probes of the gaseous environments around them, we conclude that the ICM around these FRII sources should be of low pressure, and is expected to have low X-ray luminosities.

The increase in the clustering strength around FRII sources toward high redshift is likely to be a result of the evolution of the ICM. There might be more clusters with a low-pressured ICM at high redshift, thus more radio sources can survive as FRIIs in these clusters, which is consistent with the conclusions of Hill & Lilly [4]. This interpretation is consistent with the lack of evolution of optical clusters with redshift [27] coupled with the negative evolution of the cluster X-ray luminosity function [10,12,28]. It is also consistent with the results reported by Bower *et al.* (1994)[29], who observed a sample of distant, optically selected clusters and found that the X-ray emission from these clusters is surprisingly weak compared with optically similar clusters at low redshift.

An alternative model to explain the difference in the environments and the evolution of the environments of FRI and FRII sources has been proposed by Baum *et al.* (1995) [30]. They suggest that FRII sources are associated with central black holes with high accretion and spin rates, which could be produced by mergers in low velocity dispersion systems. In this model, FRI sources are found in richer environments with higher velocity dispersions in which mergers are suppressed. The evolution of the environment of FRII sources can then be explained if there are more merger events in clusters at high-redshift. This model may be consistent with the results presented here if a higher merger rate is somehow linked to a system having a low gas pressure.

It is a pleasure to thank Lauren Jones for her work during the early stages of the project, Roger Blanford, Alan Bridle, Ed Groth, Simon Lilly, Rick Perley, and the many other scientists with whom we have discussed this work, and, especially Dan Harris, the organizer of this meeting. This work was supported in part by the U.S. National Science Foundation.